\documentclass[twocolumn,showpacs,prl,groupedaddress.letter]{revtex4}% Physical Review B

\usepackage{graphicx}% Include figure files
\usepackage{dcolumn}% Align table columns on decimal point
\usepackage{bm}% bold math
\usepackage{color}
\begin{document}

\title{Motional Coherence in Fluid Phospholipid Membranes}

\author{Maikel C.~Rheinst\"adter$^{1,2}$}\email{RheinstadterM@missouri.edu}
%\thanks{\emph{Present address:} Department of Physics and Astronomy, University of
%Missouri-Columbia, Columbia, MO 65211, U.S.A.}
\author{Jhuma Das$^{1}$}
\author{Elijah J. Flenner$^{1}$}
\author{Beate Br\"uning$^{2,3}$}
\author{Tilo Seydel$^2$}
\author{Ioan Kosztin$^{1}$}

\affiliation{$^1$Department of Physics and Astronomy, University of
Missouri-Columbia, Columbia, MO 65211, U.S.A.}

\affiliation{$^2$Institut Laue-Langevin, 6 rue Jules Horowitz, B.P.
156, 38042 Grenoble Cedex 9, France}

\affiliation{$^3$Institut f\"{u}r R\"{o}ntgenphysik, Friedrich-Hund
Platz 1, 37077 G\"{o}ttingen, Germany}

\date{\today}% It is always \today, today,

%------------------------------------------------------------------------------------
\begin{abstract}
We report a high energy-resolution neutron backscattering study,
combined with in-situ diffraction, to investigate slow molecular
motions on nanosecond time scales in the fluid phase of phospholipid
bilayers of 1,2-dimyristoyl-sn-glycero-3-phoshatidylcholine (DMPC)
and DMPC/40\% cholesterol (wt/wt). A cooperative structural
relaxation process was observed. From the in-plane scattering vector
dependence of the relaxation rates in hydrogenated and deuterated
samples, combined with results from a 0.1~$\mu$s long all atom
molecular dynamics simulation, it is concluded that correlated
dynamics in lipid membranes occurs over several lipid distances,
spanning a time interval from pico- to
nanoseconds.

%As a possible implication, the information about a
%local structural perturbance of the lipid structure could propagate
%over the bilayer, which might be relevant for the understanding of
%processes and functions involving collective structural changes.
\end{abstract}

\pacs{87.16.dj, 87.14.Cc, 83.85.Hf, 87.15.ap, 83.10.Mj}
% 87.16.dj Dynamics and fluctuations
% 87.14.Cc Lipids
% 83.85.Hf X-ray and neutron scattering
% 83.10.Mj Molecular dynamics, Brownian dynamics
% 87.15.ap Molecular dynamics simulation

\maketitle

%------------------------------------------------------------------------------------
It is speculated that atomic and molecular motions in regions of
biomolecular systems with strong local interactions are highly
correlated on certain range of time and length scales
\cite{brooks89,Lipowsky:1995}. In proteins intra-protein
correlations are believed to be essential for their biological
functioning, such as protein folding, domain motion and
conformational changes. Very recently, inter-protein correlations in
protein crystals and also membranes have been reported from
experiment and simulation
\cite{KurkalSiebert:2008,RheinstadterPRL:2008}. Experimental and
computational effort has been invested to study collective molecular
motions in phospholipid model membranes
\cite{Bayerl:2000,RheinstaedterJAV:2006,Hub:2007,Tarek:2001} to
understand the possible impact on physiological and biological
functions of the bilayers, such as transport processes
\cite{Paula:1996}, and eventually their implication on function of
membrane-embedded proteins. While fast (picosecond) propagating
collective microscopic fluctuations in the plane of the bilayer can
be understood as sound waves \cite{Chen:2001,
RheinstaedterPRL:2004}, the slow (nanosecond) in-plane mesoscopic
fluctuations (undulations) are governed by the elasticity properties
of the bilayers \cite{RheinstaedterPRL:2006}.

We studied dynamical modes at nearest neighbor distances of
the lipid molecules using the neutron backscattering technique
%The neutron backscattering technique allowed to study dynamical
%modes at nearest neighbor distances of lipid molecules
\cite{RheinstaedterPRE:2007}. These modes are too fast to be
accessed by x-ray photon correlation spectroscopy and the lateral
length scales involved are too small to be resolved by dynamic light
scattering  or the neutron spin-echo technique. Selective
deuteration was used to discriminate relaxations due to collective
molecular motions from relaxations arising from localized, single
molecule excitations. In this Letter, we examine results of
inelastic neutron scattering experiments that demonstrate the
existence of slow coherent motion of lipid molecules in the fluid
phase of phospholipid bilayers. From the in-plane scattering vector
dependence ($q_{||}$) of the measured relaxation rates, combined
with results of a 0.1~$\mu$s long all atom molecular dynamics (MD)
simulation, we find that the cooperative structural dynamics in
lipid membranes occurs over several lipid distances, spanning a time
interval from pico- to nanoseconds.

%%------------------------------------------------------------------------------------
%\section{Experimental\label{Experimental}}
The experiments were carried out at the cold neutron backscattering
spectrometer IN16 \cite{Frick:2001} at the Institut Laue-Langevin
(ILL) %in its standard setup with Si(111) monochromator and analyzer
%crystals corresponding to an incident and analyzed neutron energy of
%2.08~meV
with an energy resolution of about 0.9~$\mu$eV~FWHM
($\lambda$=6.27~\AA). An energy transfer of
$-15~\mu$eV$<E<+15~\mu$eV and a $q$ range of
0.43~\AA$^{-1}<q<$1.92~~\AA$^{-1}$ were scanned, accessing time
scales of 0.28~ns$<$t$<$4.6~ns and length scales of
3.2~\AA$<$d$<$14.6~\AA. A separate line of diffraction detectors
allowed to determine the bilayer structure in-situ. Protonated
(DMPC-h) and partially (acyl chain) deuterated
1,2-dimyristoyl-sn-glycero-3-phoshatidylcholine (DMPC-d54), along
with protonated cholesterol were obtained from Avanti Polar Lipids.
Highly oriented multi lamellar membrane stacks of several thousands
of lipid bilayers of DMPC and DMPC/cholesterol were prepared by
spreading a solution of typically 25~mg/ml lipid and lipid/40\%
cholesterol (wt/wt) in trifluoroethylene/chloroform (1:1) on 2''
silicon wafers, followed by subsequent drying in vacuum and
hydration from D$_2$O or H$_2$O vapor. Twenty such wafers (five for
DMPC-h) separated by small air gaps were combined and aligned with
respect to each other to create a ``sandwich sample'' consisting of
several thousands of highly oriented lipid bilayers (total mosaicity
about 0.5$^{\circ}$). The deuterated (protonated) samples had a
total mass of about 400~mg (100~mg) of lipid or lipid/cholesterol.
The samples were mounted in a hermetically sealed aluminum container
within a cryostat and hydrated from D$_2$O or H$_2$O vapor. By
aligning the bilayer normal at 135, respective 45 degrees with
respect to the incoming neutron beam, the momentum transfer could be
placed in the plane of the bilayers ($q_{||}$) or perpendicular to
the membranes ($q_z$). The lamellar spacing of the DMPC sample was
determined to $d_z$=54~\AA\ at T=30$^{\circ}$C, which corresponds to
a relative humidity of RH=99.6\% \cite{Chu:2005}. Five different
samples (referred to as S1,..., S5) were prepared:
\begin{center}
\begin{tabular}{c|l|c|c} Sample & Bilayer
& {Cholesterol Content} & Hydration\\\hline
S1 & DMPC -d54 & -- & D$_2$O\\
S2 & DMPC -d54 & -- & H$_2$O\\
S3 & DMPC -h & -- & D$_2$O\\
S4 & DMPC -h & -- & H$_2$O\\
S5 & DMPC -d54 & 40\% cholesterol & D$_2$O
\end{tabular}
\end{center}

Neutrons are scattered by the atomic nuclei, and each element
intrinsically has non zero coherent and incoherent scattering cross
sections. In experiments the different contributions can be enhanced
with respect to each other, by, e.g., using different isotopes or
selective deuteration. While in protonated samples the {\em
incoherent} scattering is dominant and the time-autocorrelation
function of individual scatterers is accessed, (partial) deuteration
emphasizes the {\em coherent} scattering and probes the pair
correlation function. By careful analysis of the scattering of all
samples S1 to S5, it was possible to emphasize between the
collective and self motions in the underlying dynamics. Sample S5
allowed us to investigate the effect of cholesterol incorporation on
the collective dynamics of the lipid backbone because protonated
cholesterol was used.

\begin{figure}[] \centering
\resizebox{0.95\columnwidth}{!}{\rotatebox{0}{\includegraphics{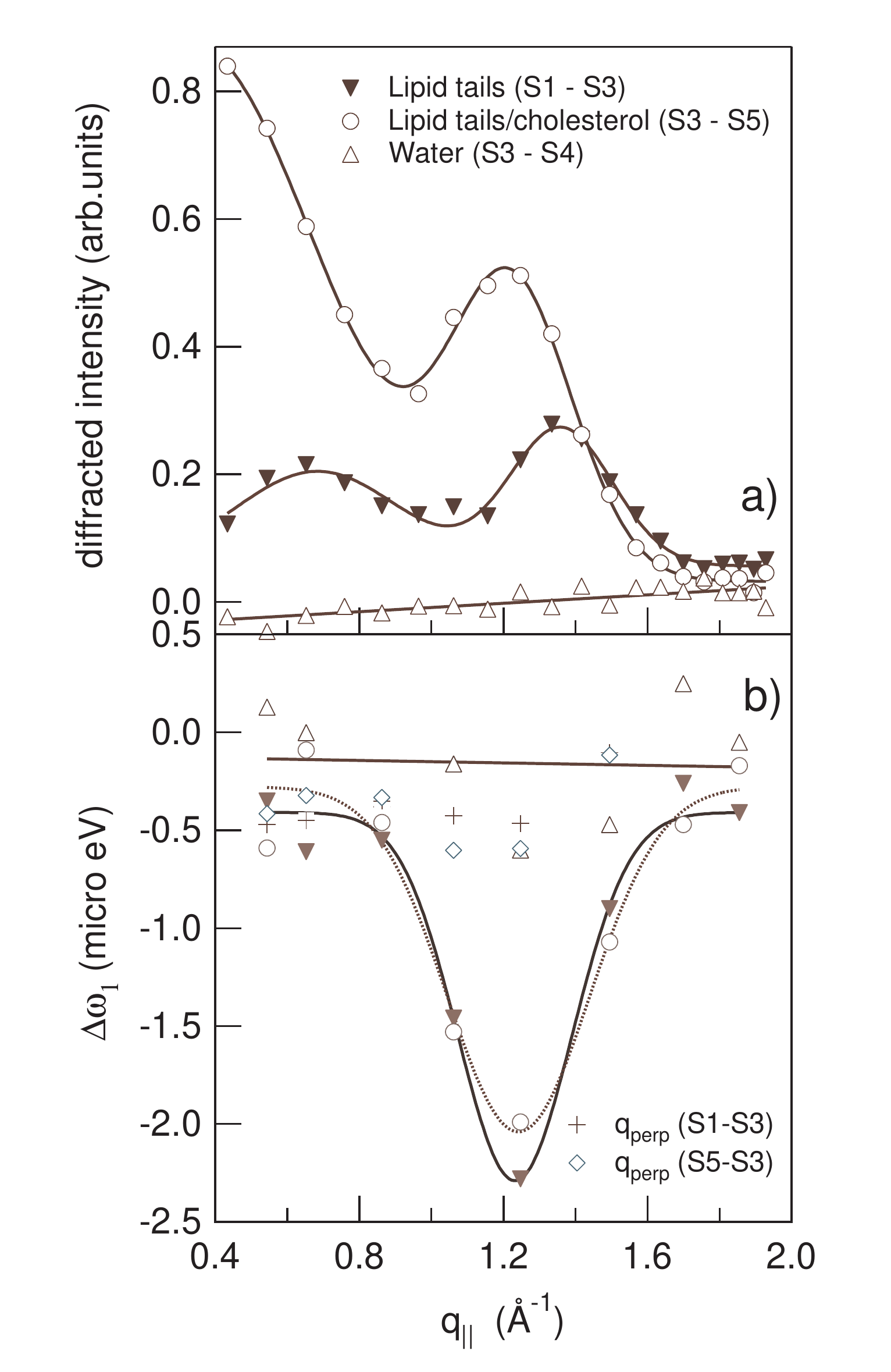}}}
%\resizebox{1.00\columnwidth}{!}{\rotatebox{0}{\includegraphics{Diffraction_difference}}}
%\resizebox{1.00\columnwidth}{!}{\rotatebox{0}{\includegraphics{w1_diff}}}
\caption[]{(Color online). (a) Diffraction patterns generated by
subtraction of
  $S(q_{||})$ for pairs of samples S1,...,S5, which emphasize contributions from collective molecular motions of the lipid
  acyl chains, and the water molecules (see text for explanation). (b) $\Delta\omega_1$ for the lipid acyl
  chains, the membrane hydration water, and the lipid chains 'distorted' by
  incorporating 40\% cholesterol. For the water, the difference
  of $\Delta\omega_1$ of samples S1 and S2 was taken.}\label{deGennes}
\end{figure}
Figure~\ref{deGennes}(a) depicts in-plane diffraction patterns,
i.e., structure factors $S(q_{||})$, at 303~K.  Subtracting
$S(q_{||})$ for samples S3 and S1 emphasizes the coherent scattering
of the lipid acyl chains.  Similarly, subtraction of $S(q_{||})$ for
S3 and S5 emphasizes the coherent lipid tail scattering in the
cholesterol sample. Finally, to emphasize coherent water scattering,
the diffraction pattern of S4 was subtracted from that of S3. There
are two peaks in $S(q_{||})$ of the lipid chains around $q_{||}=0.7$
\AA$^{-1}$ and 1.4 \AA$^{-1}$, which correspond to the nearest
neighbor distances of phospholipid head groups and the acyl chains
in DMPC, respectively.  The peaks shift to smaller $q_{||}$ values
in the presence of cholesterol, indicating a slight increase in the
nearest neighbor distances. The water scattering is enhanced towards
the water correlation peak around 2~\AA$^{-1}$, but there is no
pronounced diffraction peak within the instrumental resolution.
%%How about adding the note to the references?
%Note that when the sample contains neighboring protons (H) and deuterons (D), the
%positive and negative coherent scattering lengths of H and D add with their
%inverse signs. Therefore, the addition of D$_2$O to the protonated lipid head
%groups reduces the total coherent scattering cross section of the sample.

\begin{figure}[] \centering
\resizebox{0.90\columnwidth}{!}{\rotatebox{0}{\includegraphics{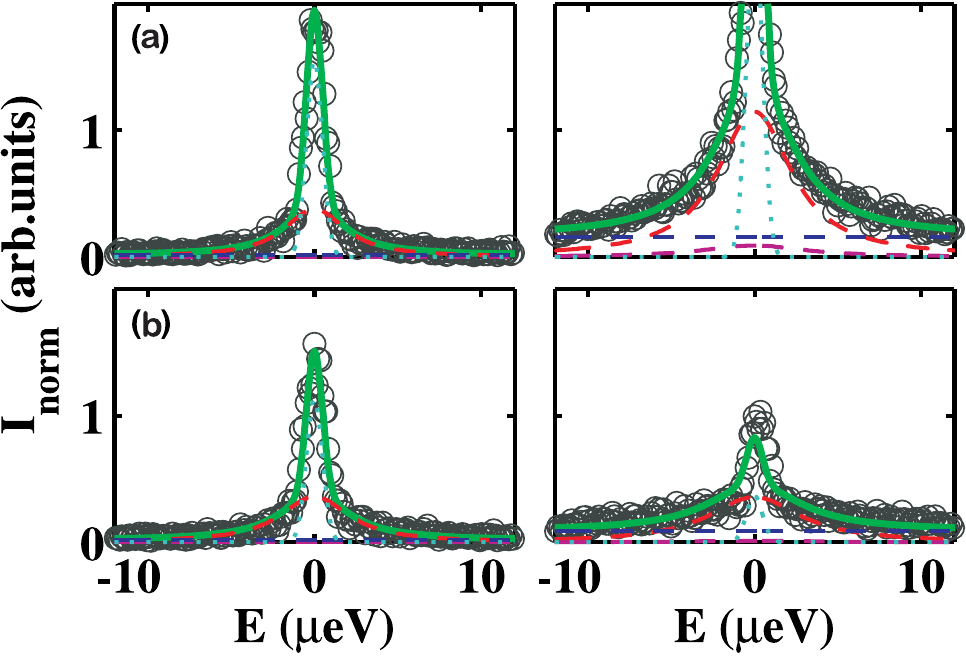}}}
%\resizebox{1.00\columnwidth}{!}{\rotatebox{0}{\includegraphics{tau1Graph}}}
\caption[]{(Color online). Exemplary inelastic spectra in the fluid
phase of the bilayers. Left column: $q_{||}=0.5$ \AA$^{-1}$; right
column: $q_{||}=1.5$ \AA$^{-1}$. a) DMPC-d54/D$_2$O (Sample S1), b)
DMPC-h/H$_2$O (Sample S4). Data have been fitted using the
instrumental resolution (dotted), two Lorentzian peak shapes
(dashed), and a constant background (dashed). The total fit is shown
as solid line.}\label{spectra}
\end{figure}
Figure~\ref{spectra} exemplarily depicts spectra measured at
$q_{||}=0.5$ \AA$^{-1}$ and 1.5 \AA$^{-1}$ for S1 and S4 at a
temperature of T=303~K, in the fluid phase of the bilayers. A fast
and a slow relaxation process were visible in the data and data were
therefore fitted using two Lorentzians and convoluted with the
instrumental resolution. Because the faster process (the broad
Lorentzian) could not be determined with sufficient accuracy, this
Letter focuses on the slower relaxation process ($\omega_1$).
$\omega_1$ might contain contributions from incoherent and coherent
scattering. A quantitative determination of the coherent and
incoherent relaxation rates would require purely coherent and purely
incoherent scatterers or neutron polarization analysis (which has
not yet been realized on a neutron backscattering spectrometer) to
discriminate the different contributions to $\omega_1$. A
convolution of incoherent and coherent relaxation processes, each
described by a Lorentzian peak shape (implicating a single
exponential relaxation process), was assumed. Since the convolution
of two Lorentzians with widths $\Delta\omega_1'$ and
$\Delta\omega_1''$ is again a Lorentzian peak shape with a HWHM of
$\Delta\omega_1=\Delta\omega_1'+\Delta\omega_1''$, by subtracting
the Lorentzian widths of spectra of protonated (mainly incoherent
scatterers) from the (partially) deuterated samples (mainly coherent
scatterers), it was qualitatively possible to distinguish the
cooperative dynamics relative from the single-particle dynamics.

Fig.~\ref{deGennes}(b) shows the difference in the quasielastic
width $\Delta\omega_1(q_{||})$ for the slow process for the lipid
acyl chains, the membrane hydration water, and the lipid chains
'distorted' by incorporating 40\% cholesterol.  So does, e.g.,
subtracting the widths of spectra S1-S3 emphasize the coherent lipid
tail dynamics, analogous to the diffraction discussed above.
$\Delta\omega_1(q_{||})$ shows a pronounced minimum around
$q_{||}=$1.22~\AA$^{-1}$ for the lipids and lipids/cholesterol
sample and qualitatively reproduces the behavior of $S(q_{||})$ in
Fig.~\ref{deGennes}(a). This points to a strong coupling of
dynamical properties to the static structure of the system and to a
softening of the coherent relaxation rate around the maximum of the
static structure factor $S(q_{||})$.  The slowing down of the decay
of the density autocorrelation function for distances corresponding
to nearest neighbors is known as {\em de Gennes narrowing}
\cite{DeGennes:1959}. The momentum dependence of the dynamical
parameters in the collective dynamics case is not trivial as in the
single-particle case but points to coherent structural relaxation.
Because the out-of-plane component ($q_{\perp}$) of the scattering
vector (also in Fig.~\ref{deGennes}(b)) shows no minimum, it is
concluded that the relaxation process is confined to the plane of
the membranes.

To gain further insight in the above relaxation process, we
performed an all atom MD simulation of DMPC and analyzed the motion
of the carbons within the lipid tails. The fully solvated system,
built from a pre-equilibrated DMPC bilayer \cite{Gurtovenko2004},
contained 128 lipid and 2577 TIP3 water molecules. The MD simulation
was performed with NAMD-2.6 \cite{NAMD} using the CHARMM27
\cite{Feller2000} force field.  The pair interactions were turned
off smoothly from 10~\AA\ to 12~\AA, and the long-range
electrostatic interactions were calculated using the PME method
\cite{essmann95-8577}. The simulation was performed in the NVT
ensemble using periodic boundary conditions. A constant temperature
of T=303~K was maintained by employing a Langevin thermostat with a
coupling constant of 0.05 ps$^{-1}$. After proper energy
minimization and thermal equilibration, a production run of
$0.1~\mu$s was carried out on 40 CPUs of a dual core 2.8GHz Intel
Xeon EM64T cluster with a performance of ~0.2 day/ns.

To examine the structural relaxation of the lipid tails we calculated the in-plane
incoherent (self) $I_s(q_{||},t) = (1/N) \sum_n \langle e^{-i
  \vec{q}_{||}\cdot[\vec{r}_n(0) - \vec{r}_n(t)]}\rangle$ and the coherent
  $I_c(q_{||},t) = (1/N) \sum_n \sum_m \langle e^{-i \vec{q}_{||}[\vec{r}_n(0) - \vec{r}_m(t)]}
\rangle$ intermediate scattering functions. The sums are taken over
the $N$ carbon atoms in the lipid tails and $\vec{r}_n(t)$ is the
position of the $n^{th}$ atom at time $t$. The scattering vector
$\vec{q}_{||}$ is parallel to the $x-y$ plane of the lipid bilayer.
By examining $I_{s/c}(q_{\parallel},t)$ we can isolate purely
coherent and incoherent contributions to the scattering due to the
carbons within the lipid tails.

\begin{figure}
\includegraphics[width=3.0in]{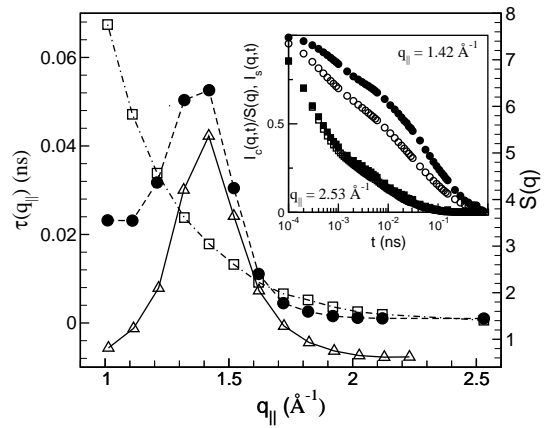}
\caption{The slow relaxation times $\tau_c$ (closed circles) and $\tau_s$ (open
  squares) as a function of wave-vector $q_{||}$. Also shown is the static
  structure factor $S(q_{||})$ (open triangles). Inset: $I_s(q,t)$ (open symbols)
  and $I_c(q,t)$ (filled symbols) for $q_{||} = $ 2.53~\AA$^{-1}$ (lower curves)
  and 1.42~\AA$^{-1}$ (upper curves).\label{sim}}
\end{figure}

The minimum in $\Delta \omega_1(q_{||})$ shown in Fig.~\ref{deGennes} suggests a
longer relaxation time for $I_c(q_\parallel,t)$ than for $I_s(q_\parallel,t)$ for
$q_\parallel$ values around the peak of $S(q_\parallel) = I_c(q_\parallel,0)$.
Shown in inset to Fig.~\ref{sim} are $I_{s/c}(q_\parallel,t)$ for $q_\parallel =$
2.5~\AA$^{-1}$ and 1.42~\AA$^{-1}$.  For $q_\parallel =$ 2.5 \AA$^{-1}$ the decay
times for $I_s(q_\parallel,t)$ and $I_c(q_\parallel,t)$ are almost identical, but
$I_c(q_\parallel,t)$ decays slower for $q_{||}$ around the first peak of the
static structure factor.  We defined the decay time $\tau(q_\parallel)$
as when $I_{s/c}(q_\parallel,\tau(q_\parallel))/I_{s/c}(q,0) = e^{-1}$.
Shown in Fig.~\ref{sim} are $\tau_s(q_\parallel)$ and
$\tau_c(q_\parallel)$ along with $S(q_\parallel)$. Note that the
nonmonotonic behavior of $\tau_c(q_{||})$ yields a minimum in the
difference $\tau_s-\tau_c$ around the peak of $S(q_\parallel)$.
Thus, the MD simulation results are in qualitative agreement with
those from the experiment and reinforces the conclusion that the
difference of the quasielastic widths $\Delta \omega_1(q_{||})$ is
due to correlated dynamics of the lipid acyl chains.

To determine the temporal and spatial extent of the correlated motion of the lipid
tails, we calculated the correlation function
\begin{eqnarray}
\label{pair}
g_{\Delta}(r,t) & = & \frac{V}{\langle\Delta{r}\rangle^2 N} \times \nonumber \\
& &  {\sum_{n,m}}^{\prime}
\langle\Delta r_n(t) \Delta r_m(t) \delta(r - |\vec{r}_n(0) - \vec{r}_m(0)|)
\rangle, \nonumber
\end{eqnarray}
where $\Delta r_n(t)=|\vec{r}_{n,||}(t)-\vec{r}_{n,||}(0)|$ represents the displacement
of carbon atoms in the $x-y$ plane of the membrane, and the prime denotes a
restricted double sum in which carbon pairs belonging to the same lipid tail are
excluded.
% Note that it is expected for displacement of atoms in the same lipid to be
% correlated.
In the absence of displacement correlations, $g_\Delta(r,t) = g_1(r) =
(V/N^2)\langle{\sum_{n,m}}^{\prime} \delta(r - |\vec{r}_n - \vec{r}_m|)\rangle$.
Thus $\Gamma(t) = \int [g_\Delta(r,t)/g_1(r) - 1]\, \mbox{d}r$ provides a measure
of the correlation of in plane displacements of the carbon atoms within the lipid
tails between different lipids \cite{Bennemann1999}.

\begin{figure}[tb]
\resizebox{1.0\columnwidth}{!}{\rotatebox{0}{\includegraphics[clip=true,
bb=0in 0in 11in 5.2in]{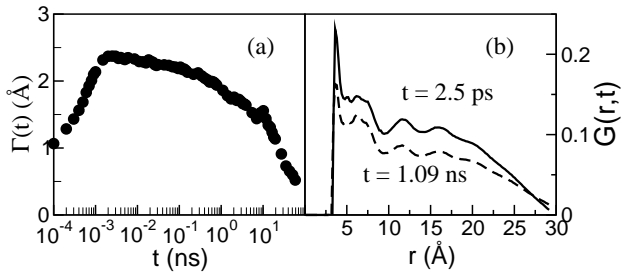}}} \caption{Measures of the
correlated dynamics $\Gamma(t) = \int G(r,t) \mbox{d}r$. $\Gamma(t)$
is a measure of the size of the correlated displacement at a time
$t$, and $G(r,t)$ measures the spatial range of the
correlations.\label{correlation}}
\end{figure}

Examination of $\Gamma(t)$, Fig.~\ref{correlation}(a), shows that the
displacements of the lipid acyl tails are correlated for times from around 1~ps to
1~ns, i.e., for much longer than the microscopic collision time but shorter than
the time needed for a lipid to diffuse one lipid diameter. This is exactly the
time window of the structural relaxation probed by the neutron scattering
experiments and corresponds to the decay times $\tau_{s/c}$.  Thus, we associate
$\Delta \omega_1$ observed in the experiments and the difference in decay times
$\tau_c-\tau_s$ to correlated motion of lipid tails.

Finally, the spatial decay of $G(r,t) = g_\Delta(r,t)/g_1(r)-1$ yields the length
scale associated with the correlated displacements. Shown in
Fig.~\ref{correlation}b is $G(r,t)$ for $t_1=$ 2.5~ps (at the peak of $\Gamma(t)$)
and $t_2=$ 1.1~ns.  For both times $G(r,t)$ does not decay to zero until around
30~\AA. Therefore, the lipid tails displacements are correlated for at least four
to six lipid diameters between $t_1$ and $t_2$.

In conclusion we found experimental evidence for a cooperative
structural relaxation process in fluid phospholipid membranes. An
all atom molecular dynamics simulations demonstrated that the
displacements of the lipid tails are correlated for up to six lipid
diameters for a time span from picoseconds to around a nanosecond.
Experimental data of the sample containing cholesterol proved that
the process is present not only in pure lipid bilayers, but also in
more complex and relevant membranes. A possible implication of this
motional coherence of the lipid acyl chains is, e.g., that
information about a local structural perturbance of the lipids could
propagate in the bilayer, which might be relevant for the
understanding of processes and functions involving collective
structural changes.
%Future
%research may include how the presence of other molecules influences
%the dynamics, and the how this correlated motion effects other
%dynamics in the liquid. Future neutron spectrometers incorporating
%polarization analysis will help to more accurately separate coherent
%and incoherent scattering signals and to quantify the effect.

We acknowledge financial support from the DFG through project SA
772/8-2. We thank T.~Salditt (G\"ottingen) for continuous support,
E.~Kats for critical reading of the manuscript and valuable
comments, and the ILL for the allocation of beam time. Computer time
was generously provided by the University of Missouri Bioinformatics
Consortium.

%------------------------------------------------------------------------------------
\bibliography{references_dmpc1,references_dmpc2}

\end{document}